# Mini-Tracker concepts for the SALT transient follow-up program

John A. Booth*[a], Michael Shara[b], Steven M. Crawford[c], Lisa A. Crause[d,e]

[a]Large Telescope Consulting Engineering, 2313 Indian Trail, Austin, TX 78703, USA
[b]American Museum of Natural History/Astrophysics, CPW & 79th St., New York, NY 10024, USA
[c]Space Telescope Science Institute, 3700 San Martin Drive, Baltimore, MD 21218, USA
[d]South African Astronomical Observatory, P.O. Box 9, Observatory 7935, Cape Town, South Africa
[e]Southern African Large Telescope, P.O. Box 9, Observatory 7935, Cape Town, South Africa

## ABSTRACT

The MeerKAT radio telescope array, the Large Synoptic Survey Telescope (LSST), and eventually the Square Kilometer Array (SKA) will usher in a remarkable new era in astronomy, with thousands of transients being discovered and transmitted to the astronomical community in near-real-time each night. Immediate spectroscopic follow-up will be critical to understanding their early-time physics – a task to which the Southern African Large Telescope (SALT) is uniquely suited, given its southerly latitude and the 14-degree-diameter uncorrected field (patrol area) of its 10-m spherical primary mirror. A new telescope configuration is envisioned, incorporating multiple "mini-trackers" that range around a much larger patrol area of 35 degrees in diameter. Each mini-tracker is equipped with a small spherical aberration corrector feeding an efficient, low resolution spectrograph to perform contemporaneous follow-up observations.

## 1. INTRODUCTION

*Cosmic Discovery*[1] offers an excellent perspective on uncovering unique new astrophysical phenomena. Each such object differs in at least one of its physical characteristics (e.g. central density, mass, luminosity) from its nearest neighbor in multi-dimensional, physical parameter-space by at least a factor of 1000. For example; red giants, main sequence stars and white dwarfs are fundamentally different phenomena, while F and G dwarfs are not. The book also makes the point that the discovery of most new classes of astrophysical phenomena demands novel instrumentation that can probe previously unexplored regions of sensitivity, time- or spectral-resolution, or wavelength space. In addition, most such instruments make their most important discoveries within just a few months or years of going online.

Surveys from the Zwicky Transient Facility[2] and GAIA[3] are already detecting an unprecedented number of transients and in South Africa, the MeerKAT[4] radio array and its optical slave, MeerLICHT[5], will soon become fully operational. These two facilities will offer the unprecedented combination of simultaneous radio and multi-band optical coverage of the southern sky, at superb sensitivity levels and spatial resolution. Having a corresponding optical image to accompany every night-time radio observation will open up the regime of simultaneous, short time-scale radio-optical correlations in astrophysical transients: dwarf novae, novae, X-ray binaries, pulsars, fast radio bursts, supernovae, gamma-ray bursts, active galactic nuclei, gravitational wave events and sources yet unknown.

The next generation of astronomical facilities will provide even larger scale monitoring of extensive regions of the sky to extremely faint limits, over a range of wavelengths and timescales. These future projects include, among others, the Large Synoptic Survey Telescope (LSST)[6], the Wide Field Infrared Survey Telescope (WFIRST)[7], the Cherenkov Telescope Array (CTA), and the Square Kilometer Array (SKA). The resulting surveys will discover vast numbers of transient objects, a small fraction of which may be representative of new phenomena. Low-resolution identification-spectroscopy is necessary to classify the sources before more comprehensive follow-up can be done on the most interesting objects.

A key challenge for the next decade is that the astronomical community is facing an overwhelming "impedance mismatch"; these transient surveys will produce a far greater number of transient sources than can be spectroscopically followed up by existing telescopes. Even after the number of sources has been whittled down by smart algorithms that identify the most interesting targets requiring immediate follow-up, the sheer number of remaining objects will still be significantly greater than the current planet-wide spectroscopic follow-up capability.  Furthermore, the limited fields-of-view of modern 10-m-class telescopes are not well suited to following up a large number of these sources, due in part to the relatively low space density of transient objects.

*booth@astro.as.utexas.edu

Telescopes with spherical primary mirrors, however, can deliver an extremely large field-of-view.  This has historically been the province of Schmidt telescopes.  For example, the 1.2-m UK Schmidt telescope has a field-of-view of 43.6 deg$^2$ and will next be used for the Taipan survey, a wide field extragalactic spectroscopic survey[8].  For telescopes with a larger primary mirror, the Schmidt design with a single, large optic for correcting spherical aberration at the primary mirror center-of-curvature is impractical.  The largest field-of-view for a current instrument on a 10-m-class telescope is 1.77 deg$^2$ for the Subaru Hyper Suprime-Cam[9].  However, spherical primary mirrors have also been used to build cost-effective 10-m-class telescopes such as the Hobby-Eberly Telescope (HET)[10] and SALT[11-13], through the use of "field partitioning"[14,15].  In the current design configuration of SALT, all of its instruments are fed by a single spherical aberration corrector mounted on a large prime-focus tracker.  The corrector has an instantaneous field-of-view (partitioned field) of about 50 arcmin$^2$, but the telescope can patrol an area of about 154 deg$^2$ (14 degrees in diameter) by moving the tracker and corrector to a different location within this area while maintaining the same telescope azimuth.

We propose to increase this patrol area by a factor of six (35 degrees in diameter), and to populate it with four to six positionable mini-trackers.  In this paper, we present design concepts for such a system that could potentially return spectra of 10,000-to-15,000 unique transients per year (four to six mini-trackers, each observing ten objects/night, for 250 clear nights/year), without significantly disrupting the telescope's present mode of operation.

### 1.1. One Example: LSST Transient Alerts

As an example, LSST is expected to produce ~10,000 alerts per minute from a 9.6 deg$^2$ area of sky.  The telescope will typically have about 1000 pointings per night, and these alerts will occur for any object that varies with respect to a previous pointing of the telescope.  This will result in tens of millions of alerts per night[4].  The vast majority of these alerts will not need immediate follow-up since many will be known variable stars, previously-discovered solar system objects, etc.  However, if even 1% of these objects appear to be interesting, this will still result in perhaps 100,000 objects per night that will benefit from spectroscopic follow-up.  These sources will have a space density of approximately 10.5 sources per deg$^2$.

## 2. BACKGROUND

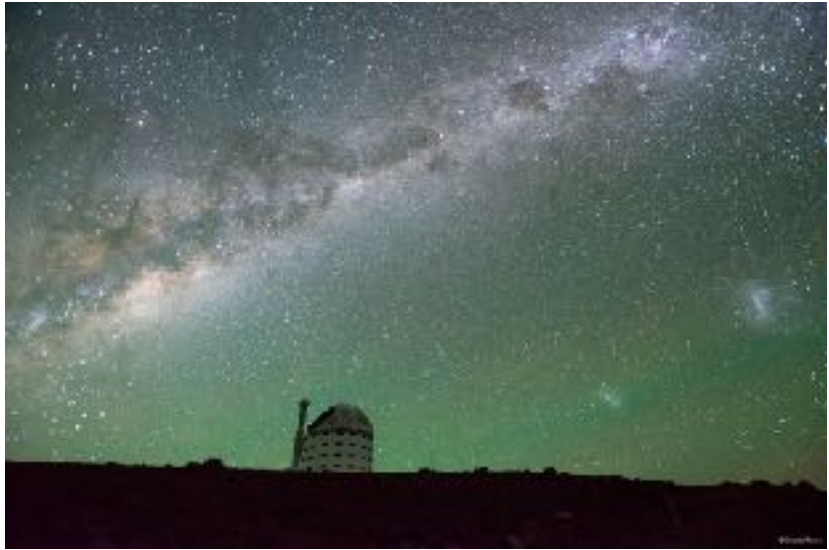

Figure 1.  SALT, the Milky Way, and the Magellanic Clouds (along with considerable airglow).  The distance between the Large and Small Magellanic Clouds, the fuzzy patches in the lower right corner of the image, is about 20 degrees, or approximately 1.5 times the current patrol area of SALT.

## 2.1. SALT description and operation

SALT is a 10-m telescope located near Sutherland, South Africa, at 32.4°S latitude, 20.8°E longitude. In an economical design adopted from the HET, it has a segmented, spherical primary mirror with 91 hexagonal 1-m segments mounted at a fixed 37 degree zenith angle. The tracker is supported by the telescope structure at a distance of 13 meters above the primary mirror. The tracker holds the spherical aberration corrector (SAC)[16] that delivers an 8-arcmin diameter field-of-view to feed a suite of instruments. The tracker follows an object across the sky during an observation (see Fig. 2), but due to its fixed elevation, there is a limited amount of time that SALT can observe any object. This track time can range from 45 minutes to about four hours, depending on the object's declination. The current instrument suite includes SALTICAM (an acquisition and imaging camera[17]), the Robert Stobie Spectrograph (RSS, a multi-purpose medium resolution spectrograph)[18], the High Resolution Spectrograph (HRS)[19,20] and BVIT (a high time resolution imager)[21].

SALT saw first light in 2005, but severe image quality problems[22,23] and RSS throughput issues led to an extended trouble-shooting, repair and commissioning phase. The telescope began full science operations in 2011 and its scientific productivity has increased steadily since then (with ~200 refereed papers in total as of mid-2018, 49 of which were published during 2017). These outputs, combined with the telescope's extremely low operating costs make SALT the most cost-effective 10-m class telescope[24]. As an entirely queue-scheduled observatory with its full complement of instruments available at all times, and with the option of breaking into the scheduled observing queue at any time with a high-priority target of opportunity, SALT is extremely well suited to observations of transients[17,25-29].

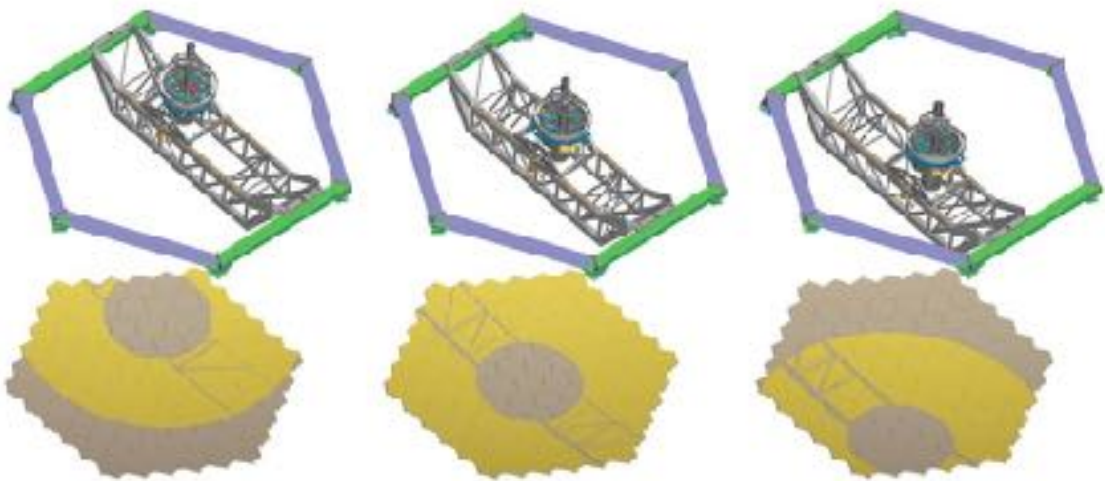

Figure 2. Tracking and effective mirror illumination for a typical SALT observation.

## 3.  SALT AND MINI-TRACKERS

This paper describes a means by which SALT could be augmented and used to provide spectroscopic follow-up of thousands of targets per year, even while executing its normal observing queue.

Observing large numbers of objects simultaneously with a SALT-style telescope will require the following three problems to be addressed:

   1) Moving targets, since the SALT primary mirror is stationary during observations

   2) Massive spherical aberration, since its primary mirror is spherical

   3) Having far more targets in the primary mirror field than the existing SALT corrector, or any practical single corrector and detector system, can possibly acquire and track

SALT, and its predecessor in the Northern Hemisphere, the HET, have solved the first two problems; using a tracking secondary system to follow the target image as it moves along the focal sphere[31,32], and having a SAC to address the second problem. The SALT SAC has an 8 arcmin field-of-view, and HET has recently upgraded its original corrector to a 22 arcmin version for the HETDEX project[33].

We propose to solve the third problem, that of large numbers of interesting targets in the field-of-view of the primary mirror, by attaching a number of small mini-trackers to the main tracker and using all or most of the main tracker's precision motion axes (depending on the concept) to track various objects simultaneously.

## 4. MINI-TRACKER CONCEPT

### 4.1. Field partitioning

The idea of correcting small portions of the large uncorrected field-of-view of a spherical primary mirror is not new. Meinel[15] and Burge[16] have described methods of "field partitioning" in cases where the fabrication and mounting of a Schmidt-type corrector plate is impractical (typically, in the cases of large telescopes.)

Both of the above concepts are predicated upon more traditional, fully-steerable telescope designs. In the case of a fixed elevation telescope like SALT, which can only rotate in azimuth between observations, one must trade aperture for field-of-view and for track length. Still, at the edge of SALT's mini-tracker-extended 35 degree diameter field-of-view, the mini-SACs see the equivalent of a 3.5-m telescope aperture.

### 4.2. What is a "mini-tracker"?

In SALT's case, the SAC partitions and corrects a relatively small (8 arcmin) field-of-view and can track this field over a range of about 14 degrees of sky. The notion behind the mini-trackers (MT) is that if one could deploy small SACs from the tracker around an expanded field (up to 35 degrees), each with its own "patrol area", one could potentially then acquire and observe several objects within 10-to-12 degrees of the main SAC boresight of the telescope. In effect, the small SACs could be observing these extra objects "for free"; that is, these observations could be achieved without materially affecting the original observation being performed by the main SAC.

In order to accomplish these observations, each MT would require the following elements:

> 1) A mini-spherical aberration corrector (MSAC) to bring the highly aberrated image in the prime focus "focal sphere" to a well-corrected, focused image. The MSAC design could be a four-mirror axial design (perhaps a small version of the existing corrector), or an off-axis, two-mirror design[19]. The field-of-view for the MSAC would need to be large enough to normally include a guide star.
>
> 2) An acquisition and guiding (AGM) module to acquire and center objects onto an optical fiber bundle, and then maintain focus and tip/tilt alignment.
>
> 3) An integral field unit (IFU), consisting of an optical fiber bundle with end treatments appropriate to the final design. This might include a lenslet array or field lens at the input end of the bundle, and a transitioning optical element at the spectrograph end.
>
> 4) A low-resolution spectrograph, or part of a spectrograph, suitable for the science objectives of the program.
>
> 5) A mechanical deployment mechanism.

Items 1) through 4) above will be largely identical for each MT. Item 5) is the main subject of this study and is treated in section 5, below.

### 4.3. Mini-spherical aberration corrector (MSAC)

A much smaller version of the current SALT SAC is envisioned as the baseline corrector for the MT heads. This corrector would be approximately 300 mm in diameter and 600 mm long, with an output beam of approximately F/3.5 and a field of approximately 1 arcmin in diameter. Depending on the results of a primary mirror field vs. vignetting study, it could have an effective aperture when fully filled of between 7-and 10-m. The four reflective elements would be fabricated from either diamond-turned aluminum and overcoated for lowest cost, or else figured from low-expansion optical glass blanks.

As an example, a small two-mirror, diamond-turned aluminum SAC called the "Surrogate SAC" was designed[34] and built in the mid-1990s and successfully used in early shakedown and testing of the HET. The Surrogate SAC is 355 mm in diameter, 472 mm long, and weighs 25 kg. It has an entrance aperture of 9.2-meters and an output focal ratio of F/1.8 (see Fig. 3).

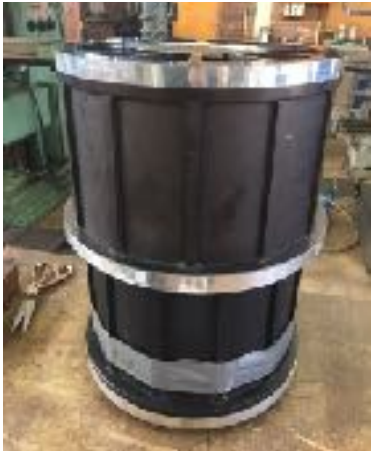

Figure 3. The "Surrogate SAC" used on the HET in the late 1990s.

### 4.4. Acquisition and guiding module (AGM)

The AGM is comprised of an optical and a mechanical subsystem. The optical subsystem will consist of a simple imaging camera, wavefront sensor, and associated optics that can acquire and center target objects on the IFU bundle. The wavefront sensor will sense focus and local tip/tilt errors of the MT. The mechanical alignment subsystem will consist of a small precision rotation stage to which the SAC is mounted, and a small commercial hexapod assembly on which the rotation stage is mounted.

The rotation stage will de-rotate the sky, and the hexapod will receive processed commands from the optical subsystem and maintain focus and tip/tilt alignment. The hexapod will also receive very small Right Ascension (RA) and declination (dec) guiding commands from the optical subsystem to maintain the object's position on the IFU. Since the hexapod can be programmed to rotate about virtually any point in space, the rotation point will most likely be about prime focus, which will eliminate "crosstalk" between tip/tilt and guiding commands.

### 4.5. IFU bundle

An IFU consisting of optical fibers will transmit the target object's light from the MT "head", along the deployment mechanism, and back to a small spectrograph mounted on the tracker payload within the central obscuration of the main corrector. The HETDEX project at McDonald Observatory now has a great deal of experience producing and safely deploying armored fiber bundles from a moving corrector to fixed spectrographs on the HET, and it is anticipated that this experience can be used to good advantage for the MTs.

### 4.6. Spectrograph

A small, low-resolution (R=1000) spectrograph, such as the one under development by SALT called MaxE[35], would be ideal to produce spectra of target objects from each MT. A small number of such spectrographs rack-mounted in an enclosure on the SALT payload assembly would minimize fiber length and maximize the amount of light reaching each instrument.

## 5. DEPLOYMENT MECHANISM CONCEPTS

### 5.1. General

There are a number of methods that can be used to deploy the MT heads described in the previous section, each having advantages and disadvantages. We have considered more than a dozen different approaches, and have narrowed the field down to what we believe are the five most practical concepts. Of the five concepts described in this paper, the first two concepts (PMRA, or Payload-mounted Robot Arm, and PMXY, or Payload-mounted X-Y Stage) are nearly identical to the second two concepts (CMRA, or Carriage-mounted Robot Arm, and CMXY, or Carriage-mounted X-Y Stage), differing only in their mounting points.

The fifth concept deploys from two to four additional "Y" rails off the main tracker beam, and the MT heads are supported and positioned by a substage on these rails.

Note: For the purposes of this concept study, "tracking" refers to movement in tracker X, Y, tip, and tilt. Sky de-rotation is accomplished open-loop using individual rotation stages in each MT head, and focus is sensed and maintained locally by the MT head.

Each variant has a slightly different "patrol area", or section of the entire SALT field that is accessible at any given time, given the main tracker position.

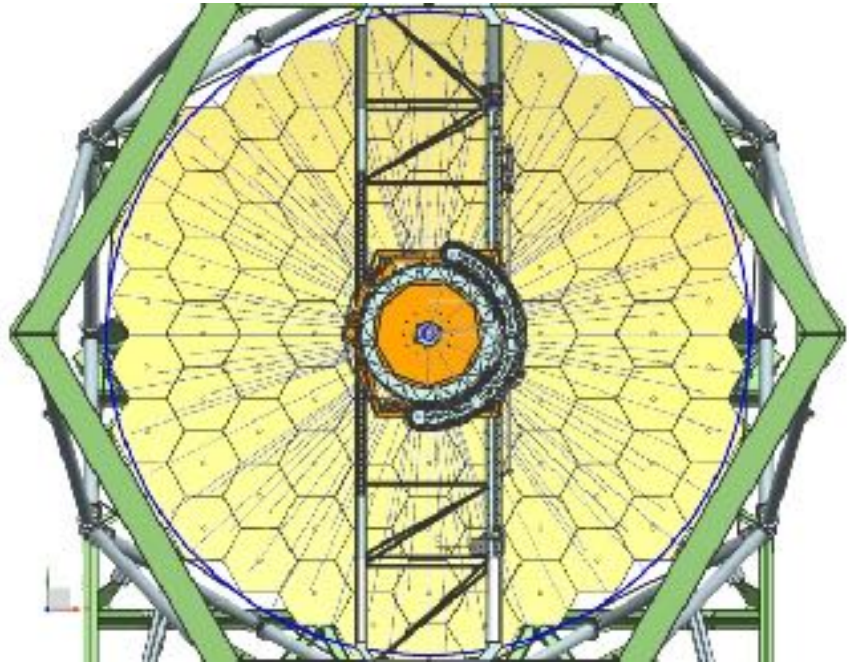

Figure 4. View "down the barrel" of SALT, showing the 91-hexagonal segment primary mirror array and the existing tracker structure and payload at center (spanning the upper structural hexagon from top to bottom in the figure). Light rays are shown as lines radiating inward from the primary mirror to the payload.

### 5.2. Concept PMRA – Payload-mounted Robot Arm

Robotic swing arms mount to the Non-Rotating Structure (NRS) of the Tracker Payload, and position the MT heads on the focal sphere, on desired targets. No additional tracking is required since the MT is effectively mounted to the upper end of the tracker hexapod. Only small guiding and tip/tilt/focus corrections are necessary, accomplished autonomously by the MT control system.

An important advantage of this concept is simplicity of operation. The MT head can be moved to a target on the focal sphere and the positioner fixed in place for the duration of an exposure. The MT head will acquire and center the target, and using its imager, wavefront sensor, and small hexapod, will guide in RA and declination, and maintain focus.

Shown below in Figs 5 and 6 are schematic diagrams of a "R-Theta" type mechanism, in which the arm is pivoted from a rotating base mounted to the payload (Theta motion), and a single axis stage (R or radial motion) is translated to position the MT head to the target. The patrol area for each arm is about 16 deg$^2$.

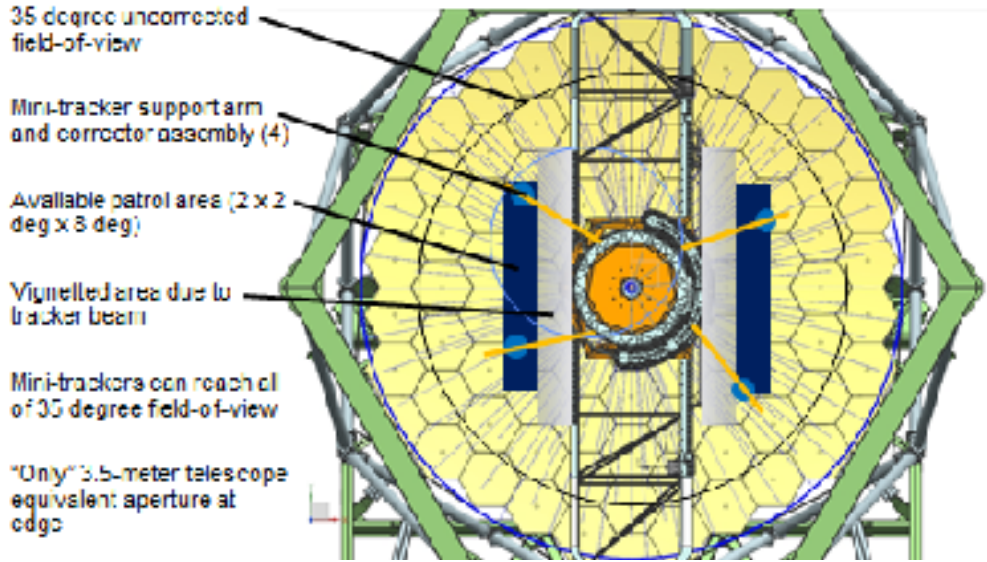

Figure 5. Concept PMRA - Payload-mounted Robot Arm, Top View. Note that the robot arms extending radially from the Payload center can swing through about 270 degrees of rotation.

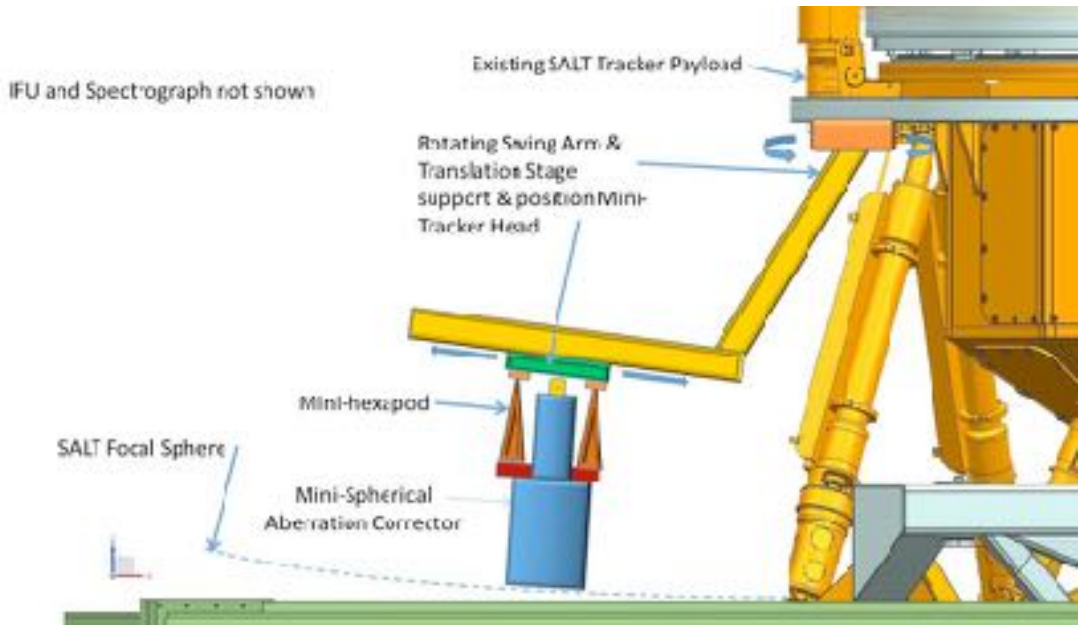

Figure 6. Concept PMRA - Payload-mounted Robot Arm, Side View.

### 5.3. Concept PMXY – Payload-mounted X-Y Stage

A large, "thin" rectangular X-Y gantry stage mounts to the NRS on fixed struts above the level of the focal sphere (see Figs 7 and 8). MT heads are positioned around the patrol areas using the gantry stages and fixed in place on each target. As in Concept PMRA above, no additional tracking is required, only small guiding and tip/tilt/focus corrections are necessary, accomplished autonomously by the MT control system.

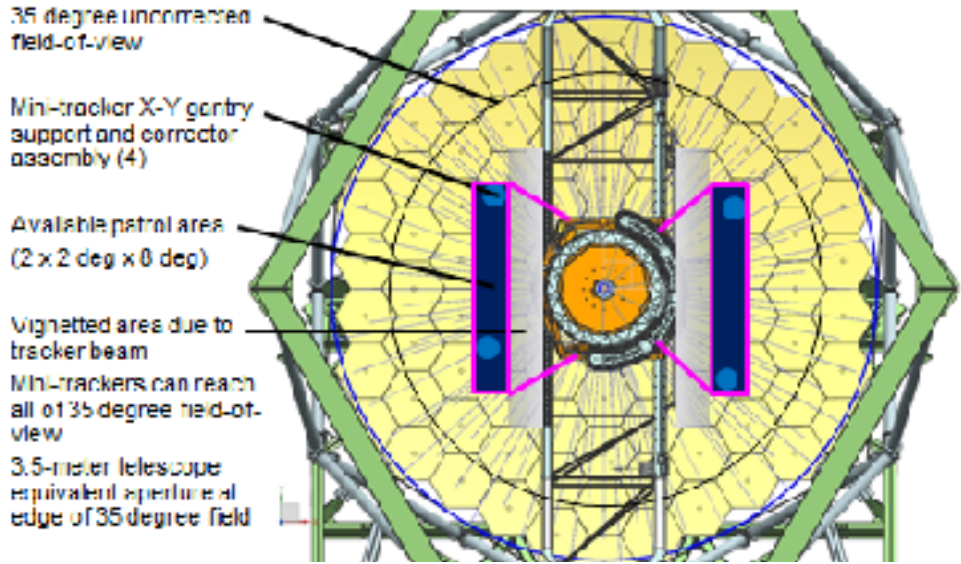

Figure 7. Concept PMXY Payload-mounted X-Y stage. Mini-correctors can roam about the available patrol areas on small gantry stages. (The side view of this concept is nearly identical to Fig. 6, where the swing arms become support struts for the X-Y gantry stage.)

### 5.4. Concept CMRA – Carriage-mounted Robot Arm

Since they are mounted on the carriage rather than the payload, both this concept and Concept CMXY – Carriage-mounted X-Y Stage below require tip/tilt tracking to keep the MT head normal to the focal sphere. They also require tracking in focus, and the depth of the focal sphere at the corner of the proposed patrol area, 12 degrees from the tracker center of travel, is about 290 mm. This is beyond the range of a typical commercially-available hexapod, and a custom-built hexapod or separate Z stage may be required.

Mounting the device on the tracker carriage appears to be simpler and more straight-forward mechanically than mounting it on the tilting part of the Payload, and a more detailed mechanical study is required to make this trade.

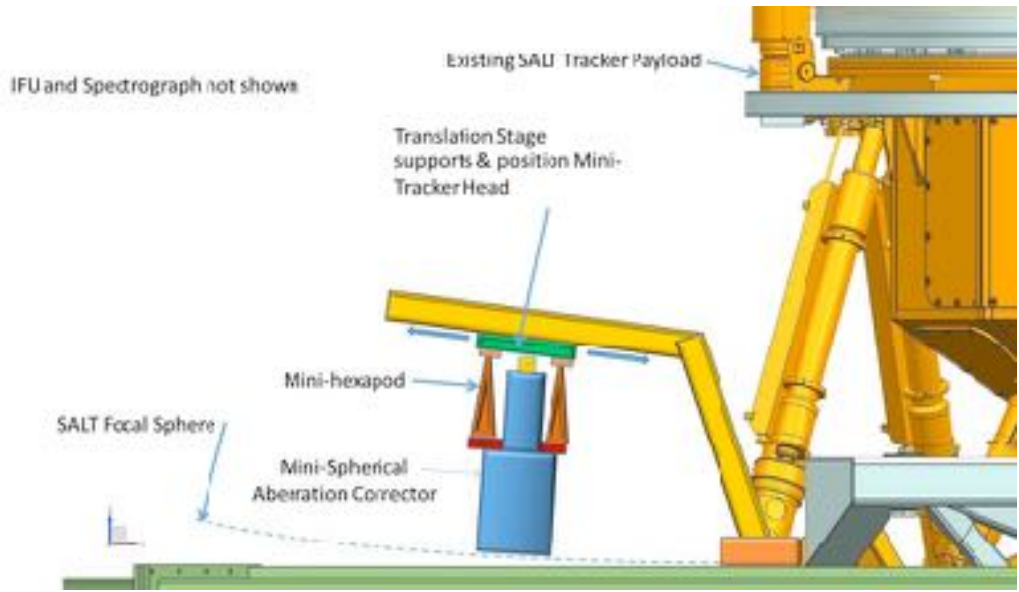

Figure 8. Concept CMRA – Carriage-mounted Robot Arm. The robot arm is mounted to the carriage as shown, and swings through similar arcs to those shown in Fig. 5.

### 5.5. Concept CMXY – Carriage-mounted X-Y Stage

Similar to Concept PMXY but mounted to the tracker carriage rather than the payload, this concept provides a large gantry stage to position the MT head, but the head must include tip/tilt tracking motion and a large tracking focus range as with Concept CMRA, above.  Again, either a longer travel hexapod (300 mm plus tip/tilt simultaneously) or a separate Z stage is required.  Schematic diagrams of this concept are essentially identical to Fig. 5, the top view of the payload-mounted X-Y stage, and Fig. 8, the Carriage-mounted robot arm side view, where the robot arm shown in Fig. 8 just becomes a set of static support struts for the X-Y stage.

### 5.6. Concept TBXY – Tracker Beam-mounted X-Y Stage

Concept TBXY (not illustrated) mounts a smaller Y-beam on a parallelogram mechanism on either side of the main Y tracker beam.  An MT head is mounted on a small Y-carriage which travels up and down the smaller Y-beam.  Positioning in the X direction is accomplished with the parallelogram deployment mechanism, and this is fixed relative to the tracker beam once the MT head is positioned on a target.  This would allow similar coverage to the two 2 x 8 degree dark rectangles shown in Figs 5 and 7, but would require a driven Y-stage for each MT, along with the tip-tilt and large focus requirements of the carriage-mounted concepts.

This concept would only be considered if insurmountable difficulties were encountered with each of the first four concepts described above.

## 6. DISCUSSION OF DEPLOYMENT CONCEPTS

### 6.1. Field coverage vs. mirror illumination for an MT

It is important to note that all of the above MT deployment concepts must work around the existing tracker, which is of course optimized to take advantage of the "fattest" part of the primary mirror during a typical observing track, as shown in Fig. 2.  Assuming for the moment a 10-m aperture MT SAC, the closest an MT can be positioned to the existing SAC is about 8.5 degrees without vignetting from the lower structural members of the main tracker beam (see Fig. 9).

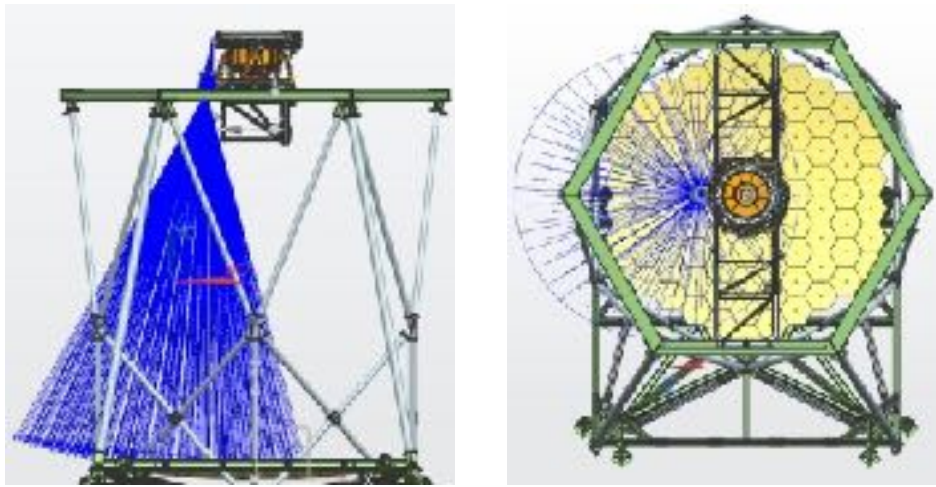

Figure 9.  Nearest position of an MT SAC to the tracker beam, front and top views.

Let us look at a typical example: If the tracker starts a full track from the extreme left position, for instance (7 degrees from the above centered position), the MT starts the track at 15.5 degrees, and sees only about 18 primary mirror segments initially.  This is the equivalent of a 4.7-m aperture Ritchey-Chrétien telescope, and things do not get better very fast for the MT, owing to the presumed ~4-m diameter central obscuration of the MSAC.  This obscuration, mapped to the primary mirror, begins to appear on the primary soon after the beginning of the track.  For more outboard patrol positions of the MT, the illumination becomes even worse, diminishing down to virtually no light at the extreme corners at the beginning or end of certain tracks (these positions become useful as the tracker approaches the center.)  Hence, target selection, based on brightness, and taking the total effective aperture into account, will be critical to the successful operation of the MTs.

### 6.2. Optimal Deployment Concept

Concept PMRA, the Payload-mounted Robot Arm, appears to be the optimal concept considered in this study, from the standpoint of simplicity of design and operation. The payload performs all of the tracking operations except sky derotation, which is easily accomplished by mounting the MT SAC on a commercially-available rotation stage. Once developed, the design can be prototyped and thoroughly tested off-telescope, then mounted on the telescope payload with minimal interference to operations for additional testing.

At this very early concept stage it appears possible to mount four such MTs to the payload, and perhaps as many as six. There are several possible configurations for the robot arm design itself (note that the concept shown only rotates from its mounting point on the payload, and has no other articulation).

Concept PMXY, the Payload-mounted X-Y stage, has similar advantages to the above concept, but is a larger assembly that may be more difficult to prototype and mount on the payload. It would likely obscure slightly more of the primary mirror that the robot arm concept, and the outer corners of the stage will be less useful from an effective aperture standpoint.

Concepts CMRA and CMXY, similar to Concepts PMRA and PMXY but mounted to the carriage, and Concept TBXY, the tracker beam-mounted concept, would only be pursued if one of the first two concepts were for some reason impractical to implement, owing to the additional complexity in tracking with these last three concepts.

## 7. NEXT STEPS

This small study of possible MT configurations and deployment concepts for SALT has been extremely limited in scope. We have attempted to lay out the most promising MT deployment concepts that could be adapted to SALT in a practical way. Clearly there is much work to be done for this capability to become a reality. An initial feasibility study that would address, at a minimum, the following issues in more detail seems to be the logical next step:

> 1) Transient science trade study to include trades of number of MTs, field, range of motion, and patrol field and track length vs. effective telescope aperture
>
> 2) MT SAC design and trade study:
>
>> - Diamond-turned vs. glass mirrors, optical coatings, cost
>> - Field size vs. SAC size, effective aperture, and mass
>> - Need for internal moving baffle or telescope structure baffling
>
> 3) Optical and mechanical design development of acquisition and guiding module and IFU bundle and routing
>
> 4) Deployment mechanism design development options, trade study:
>
>> - Mechanical stability, flexure analysis, windshake analysis
>> - Maximum feasible and affordable number of MTs
>> - Field coverage vs. primary mirror illumination, cost
>
> 5) Preliminary project plan, project cost, and schedule

## 8. CONCLUSION

Upcoming large-scale surveys associated with MeerKAT/MeerLICHT, LSST and SKA will soon begin identifying enormous numbers of transients. A substantial fraction of these will require rapid spectroscopic follow-up observations to identify the most interesting objects for more detailed study. A subset of these targets will undoubtedly represent entirely new astrophysical objects/phenomena, while others will contribute statistics to fill in gaps in our current understanding of particular types of sources and events.

Present and planned future facilities will not be able to keep pace with the alert streams predicted to accompany these surveys and so the development of new optical spectroscopic follow-up capabilities is essential. Given SALT's location in the southern hemisphere, its close proximity to MeerKAT (and the future SKA) and MeerLICHT (which is located at the same observing site as SALT), it is clear that there is a significant niche to be exploited here.

Additional work is necessary to fully develop the SALT mini-tracker concept explored in this paper. However, this initial study indicates that there may be a relatively straightforward way for SALT to take strategic advantage of its unconventional design. A set of deployable mini-trackers would allow us to capitalize on the huge uncorrected field-of-view of the telescope's primary mirror and at least quadruple the number of targets simultaneously observable by Africa's Giant Eye. If we hope to take advantage of the ground-breaking surveys on the horizon, we would need to complete a formal design study and build a working prototype mini-tracker within the next two years.

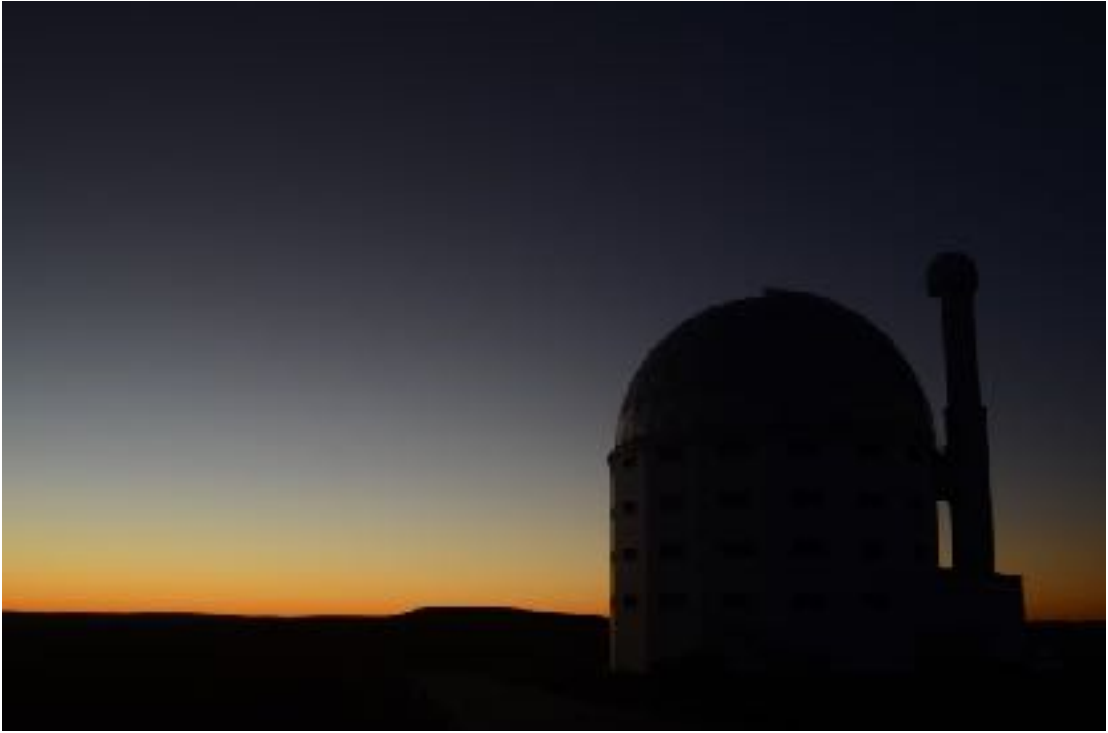

Figure 10. SALT at sunset.

## ACKNOWLEDGEMENTS

The authors wish to thank members of the SALT Operations team and the South African Astronomical Observatory who contributed to the solid modeling work for this project. In particular, the efforts of Jonathan Love, Eben Wiid, and Chantal Fourie were essential in completing this study. Ms. Fourie also provided the picture of SALT in Fig. 2. We also wish to thank John Good for his valuable insights and assistance, and Hanshin Lee for his insights on smaller SACs and the use of his illuminated aperture calculator.